# Copper underpotential deposition on boron nitride nanomesh

Stijn F. L. Mertens [1,2]*

[1] Department of Chemistry, KU Leuven, Celestijnenlaan 200F, 3001 Leuven, Belgium

[2] Institut für Angewandte Physik, Technische Universität Wien,

Wiedner Hauptstrasse 8–10/E134, 1040 Wien, Austria

mertens@iap.tuwien.ac.at; stmerten@gmail.com

**Abstract**

The boron nitride nanomesh is a corrugated monolayer of hexagonal boron nitride (h-BN) on Rh(111), which so far has been studied mostly under ultrahigh vacuum conditions. Here, we investigate how copper underpotential deposition (upd) can be used to quantify defects in the boron nitride monolayer and to assess the potential window of the nanomesh, which is important to explore its functionality under ambient and electrochemical conditions. In dilute sulfuric acid, the potential window of h-BN/Rh(111) is close to 1 volt, i.e. larger than that of the Rh substrate, and is limited by molecular hydrogen evolution on the negative and by oxidative removal on the positive side. From copper upd on pristine h-BN/Rh(111) wafer samples, we estimate a collective defect fraction on the order of 0.08–0.7% of the geometric area, which may arise from line and point defects in the h-BN layer that are created during its chemical vapour deposition. Overpotential deposition (opd) is demonstrated to have significant consequences on the defect area. We hypothesise that this non-innocent Cu electrodeposition involves intercalation originating at initial defects, causing irreversible delamination of the h-BN layer; this effect may be used for 2D material nanoengineering. On the relevant timescale, upd itself does not alter the defect area on repeated cycling; therefore, metal upd may find use as a general tool to determine the collective defect area in hybrids between 2D materials and various substrate metals.

**1.    Introduction**

Two-dimensional materials, including graphene [1] and metal–organic frameworks [2], are enjoying exceptional electrochemical research activity for a number of reasons to which non-scalability and economy of resources belong.  Wide-band-gap insulators, such as hexagonal boron nitride (h-BN), are less obvious material candidates, as—with few exceptions [3]—high electrical conductance appears paramount for electrochemistry to be possible.  For a single- or few-layer dielectric on a free-electron metal, however, the electronic wave function extends sufficiently for tunnelling across this barrier to occur [4].  In addition, the question whether an ultrathin layer of a dielectric on a conductor remains dielectric in nature is not trivial, and requires the interaction between the substrate and overlayer to be considered in detail [5].  As a recent example of such counterintuitive behaviour, electrocatalytic properties of BN nanosheets on Au have been demonstrated [6, 7].

A particularly intriguing dielectric–metal hybrid is the boron nitride nanomesh [8, 9], an atomically thin layer of h-BN on Rh(111).  Due to a 10% mismatch in lattice constants between the Rh(111) substrate and the h-BN overlayer, and suitably strong binding between the two materials [10], a pronounced corrugation with a coincidence lattice constant of 3.2 nm emerges, the details of which have been studied in considerable detail over the past decade.  The superstructure is characterised by areas with registry between the N and Rh atoms where strong binding occurs ("pores" of the nanomesh) and areas with weaker binding (in part even repulsive interactions), where the h-BN monolayer buckles up (so-called "wires").  A major consequence of the h-BN monolayer corrugation is the existence of in-plane dipole rings [11], which are responsible for trapping of polarisable atoms and molecules where the gradient of the electric field is largest, that is, at the edge of each pore. To date, all published work regarding molecular trapping on the nanomesh has been in vacuum [11-13], even though solution-based studies are underway that demonstrate the usefulness of the nanomesh as a substrate for intact immobilisation of reactive species such as monodisperse tungsten oxide clusters [14].

Recently, we demonstrated that electrochemical intercalation of hydrogen renders the boron nitride nanomesh flat [15], and that the effect of this nanoscale transformation can be seen in dynamic contact angles of an electrolyte drop, six orders of magnitude larger than the nanomesh corrugation. This observation makes the nanomesh a promising model system for the study of wetting, adhesion and stiction (static friction). Naturally, linking microscopic properties with their macroscopic expression implies that the density of defects should be kept to a minimum. Also other processes based on 2D materials, such as field-driven sieving of isotopes [16], electrochemical exfoliation to produce freestanding one-atom-thick crystals [17], and the prospect of osmotic pumps for power generation [18] require a detailed understanding of their electrochemical stability. For the boron nitride nanomesh, the stability of the nanomesh in water [9] and electrolytes [15, 19] has been demonstrated in principle, but more detailed studies are essential to explore the full potential of this class of materials.

In the years following the discovery of the boron nitride nanomesh, substantial efforts have been invested in understanding the deposition conditions that lead to highly uniform monolayer h-BN growth [20-22]. A consensus has emerged that high-quality layers require step-flow growth conditions, as the formation of line defects where h-BN domains meet is unavoidable. Suitable growth conditions include promoting fast diffusion of adsorbed precursor molecules and limitation of the number of nuclei from where growth proceeds. However, reliable yet fast and simple experimental methods to quantify the defect density remain necessary, as direct observation with a scanning probe technique is not scalable to large areas.

Here, copper upd on the h-BN nanomesh is used to quantify the collective defect area, and we propose that metal upd can be applied in general as a sensitive probe for defects in 2D materials on suitable metal substrates. Classically, hydrogen upd has been used often to determine the microscopic, as opposed to the geometric, surface area [23-25], subject to assumptions such as commensurability between upd layer and metal surface (e.g., one H atom per metal atom). For the boron nitride nanomesh, the added complexities of hydrogen intercalation, and the apparent limit at 1/3 of a monolayer [15, 26], make such presumed simple calibration impossible. Using copper electrodeposition, we explore how the upd signature changes as a function of the width of the electrochemical excursion to which we subject the nanomesh, both in negative (towards copper opd) and positive direction (where the nanomesh and/or Rh(111) substrate are oxidised).

## 2. Experimental

Boron nitride nanomesh was grown in a dedicated apparatus [27] on four-inch Si(111) wafers protected with a 40-nm yttria stabilised zirconia (YSZ) diffusion barrier and covered by a 150-nm thick single-crystalline rhodium film. Briefly, the Rh(111)-coated film was cleaned by oxygen and argon plasmas, and high-temperature degassing and annealing. The growth of h-BN itself occurred by exposing the substrate at 820 °C to 675 Langmuir (1 L = $10^{-6}$ torr·s) borazine [(HBNH)$_3$] as a precursor, which leads to self-limiting monolayer growth [8, 27]. $1 \times 1$ cm$^2$ samples were cut from these wafers in a clean room and protected by a UV-cured polymer film until seconds before use. Reference measurements on Rh(111) were carried out on a Clavilier-type rhodium single crystal (Mateck GmbH, Germany), oriented (miscut ~0.1°) and polished to mirror finish. Prior to measurements, the bead crystal was flame-annealed and cooled in a 2:1 Ar:H$_2$ atmosphere, which is known to yield good-quality (111) terraces [28].

Cyclic voltammetry was performed using a Metrohm–Autolab PGSTAT32 potentiostat. Measurements on wafer samples were carried out in a single-compartment PTFE cell pressed onto the sample by means of a 4-mm diameter Kalrez O-ring, holding a Pt wire counter and Ag/AgCl/3 M NaCl reference electrode. Rh(111) single-crystal measurements were performed in hanging-meniscus configuration in a standard two-compartment glass cell carrying a reversible hydrogen reference and Pt wire counter electrode. All measurements in this paper are reported with reference to the reversible hydrogen electrode (RHE). The electrolytes were prepared from ultrapure $H_2SO_4$ (Merck suprapur), copper sulfate (99.999% metals basis, Alfa Aesar) and water (Milli-Q, Millipore, 18.2 MOhm cm, ≤3 ppb total organic carbon). All glassware and the PTFE cell were cleaned by boiling in 20% nitric acid and rinsing with ultrapure water; the Kalrez O-ring was treated with caroic acid (3:1 conc. $H_2SO_4$:30% $H_2O_2$; *Caution! This mixture is very aggressive and should be handled with extreme care*) and rinsed with ultrapure water. Electrolytes were degassed by passing high-purity Ar (Air Liquide grade 5.0) through the cell solutions before, and over them during measurements.

Atomic force microscopy imaging was performed on a commercial Multimode AFM (Veeco), equipped with a Nanoscope IV controller and a type E scanner. Images were recorded in amplitude modulation mode, under ambient conditions, and using Silicon cantilevers (Olympus; AC160TS; resonance frequency ≈300 kHz; spring constant ≈ 20-80 pN/nm). Typical scans were recorded at 1–3 Hz line frequency, with optimized feedback parameters and at 512 × 512 pixels. For image processing and analysis, Scanning Probe Image Processor (v6.4; Image Metrology) was employed. Image processing involved background correction using global fitting with a third order polynomial, and line-by-line correction through the histogram alignment routine.

## 3. Results

The cyclic voltammogram (CV) of the nanomesh substrate alone, Rh(111), in 0.1 M $H_2SO_4$ is shown in **Figure 1a**, and is in accordance with previous measurements [23, 25]. On the negative side, hydrogen upd causes a sharp peak close to the onset of molecular hydrogen evolution. The charge under the peak, assuming a 1-to-1 ratio of adsorbing hydrogen and rhodium atoms (leading to 256 $\mu C\ cm^{-2}$ [24]), can be used to determine the microscopic surface area of the electrode, and current densities here have been calculated by this method. During the anodic scan, an oxidation feature that is ascribed to the formation of a Rh surface oxide is observed at 0.85 V [25]. The concomitant Rh oxide reduction feature during the cathodic scan is found at 0.7 V. Between hydrogen adsorption/desorption and the onset of Rh oxidation, a potential window of ca. 0.6 V is available where no Faradaic processes take place, and higher potentials should be avoided if surface oxidation is not intended.

**Figure 1b** shows the voltammetric behaviour of the same electrode in 1 mM $Cu^{2+}$ + 0.1 M $H_2SO_4$. Starting at potentials where neither Rh oxidation nor copper reduction occurs, the cathodic scan shows two peaks with a limited separation of less than 200 mV. The scan rate has a strong effect on the position of the peaks (Figure 2b, red trace), pointing at sluggish kinetics. For this reason, upd features in the following were always measured using a scan rate of 2 mV $s^{-1}$. The first cathodic peak, at 0.4 V, is assigned to upd of $Cu^{2+}$, possibly with coadsorbing sulfate [29]. The second peak, at 0.25 V, is ascribed to copper overpotential deposition (opd). Reversal of the scan direction leads to a typical stripping peak from a solid electrode, followed by an inverse Cu upd peak at 0.55 V. The charge under the cathodic and anodic upd peaks amounts to identical (within uncertainty margins) values of (410 ± 20) $\mu C\ cm^{-2}$.

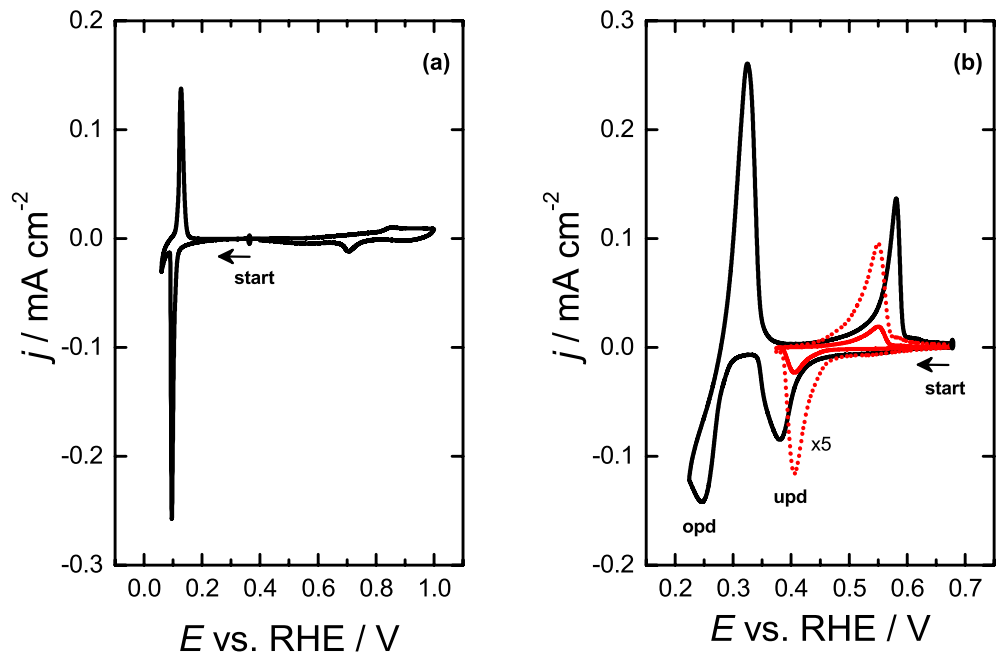

**Figure 1.** Cyclic voltammogram of Rh(111) (single crystal) in (a) 0.1 M $H_2SO_4$ and (b) 1 mM $Cu^{2+}$ + 0.1 M $H_2SO_4$. Scan rate, 10 mV $s^{-1}$; (b) upd only, 2 mV $s^{-1}$.

We now turn our attention to the electrochemical behaviour of the boron nitride nanomesh. **Figure 2** shows a representative CV of pristine h-BN/Rh(111) nanomesh in 0.1 M $H_2SO_4$. Over a potential range of close to 1 V, the only feature during the first cycle (black trace) is the signature of hydrogen upd at about 0.1 V, just before evolution of molecular hydrogen limits the potential window on the negative side. As was shown previously [15], the H-upd signal, which amounts to ca. one third of a monolayer, is related to *intercalation* of atomic hydrogen between the Rh(111) substrate and the h-BN overlayer. On moving to more positive potentials, a pronounced irreversible oxidation peak is encountered when $E > 1.1$ V. During the second cycle (red trace), the interfacial capacitance and H-upd peaks are markedly larger. Furthermore, all features typical for the bare Rh(111)|0.1M $H_2SO_4$ interface are found—in particular the oxidation of the Rh(111) at 0.8 V—as discussed previously. We ascribe the irreversible oxidation feature during the first cycle to the oxidative removal of the h-BN layer, convoluted with the oxidation of the Rh(111) outer layer, which normally occurs at less positive potentials as seen during the second cycle. Consequently, in the h-BN/Rh(111) hybrid, the boron nitride monolayer effectively passivates the substrate metal.

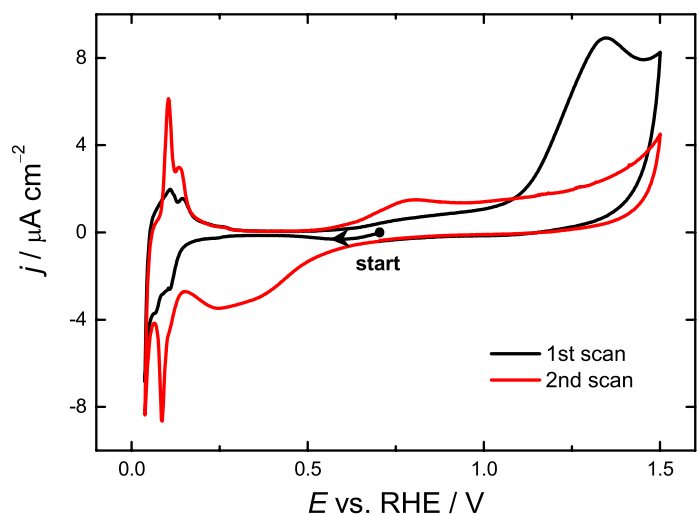

**Figure 2.** CV of h-BN/Rh(111) (wafer sample) in 0.1 M $H_2SO_4$. Scan rate, 10 mV $s^{-1}$.

The corresponding measurement on pristine h-BN/Rh(111) in the presence of copper(II) ions, and limited to the upd region of $Cu^{2+}$, is shown in **Figure 3a**, inner trace. Minimal peaks for $Cu^{2+}$ upd and the inverse process are observed, which remained unchanged on multiple cycling. This is important if the corresponding charges are to be used as a measure of defect area, because the upd process in that case must not alter the characteristics of the studied interface. Also holding the potential at 0.3 V for extended periods up to 600 s did not increase the reverse upd signal. Integration of the charge indicates an electroactive surface area of 0.7% of the geometric area. Extension of the CV to the Cu opd region is shown in **Figure 3b**. At a similar potential as on bare Rh(111) (**Figure 1b**), the opd process is observed.

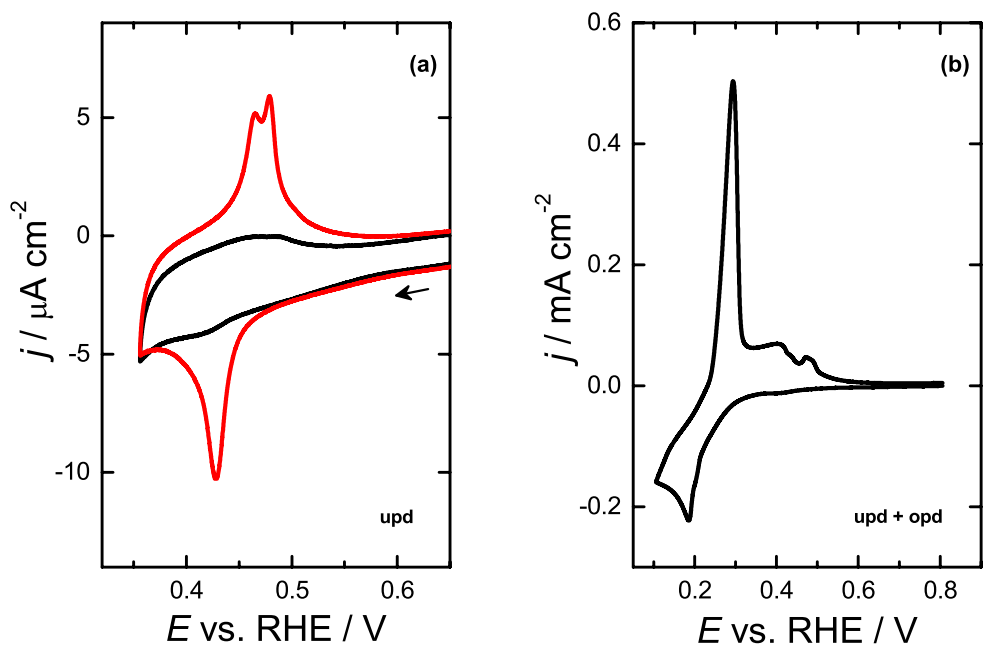

**Figure 3.** CV of h-BN/Rh(111) (pristine wafer sample) in 1 mM $Cu^{2+}$ + 0.1 M $H_2SO_4$ in the (a) Cu upd region before (inner trace) and after excursion into the opd region (outer trace) and (b) Cu upd + opd region. In these measurements, no potentials above 0.8 V were applied. Scan rate, (a) 2 mV $s^{-1}$; (b) 10 mV $s^{-1}$.

When the Cu upd region is measured again after the excursion into the opd region, a striking difference with the initial measurement is found, **Figure 3a**, outer trace: the upd peaks and their reverse have increased by a factor of 28. Multiple cycles in the upd region alone confirm the stability of the upd peaks, and, as on the pristine h-BN/Rh(111) nanomesh sample, indicate that Cu upd does not alter the characteristics of the interface on the timescale of the measurement, in stark contrast with the opd process. Comparing the anodic scan on the bare rhodium substrate with the behaviour of the nanomesh sample, the single peak on Rh(111) appears split on h-BN/Rh(111). A similar peak splitting has been observed for hydrogen intercalation on this material [15], but a source for this phenomenon has not yet been identified.

Repetition of this procedure (single scan of the opd region, multiple scans of the upd region), **Figure 4a**, shows that with every excursion into the opd region, the subsequent upd peaks increase (from 22% of the geometric area after a single excursion to 34% after four). In addition, the extent of the opd excursion affects the degree of change in upd afterwards, with less Cu deposited leading to a lesser change post-opd (data not shown).

Close inspection of the stripping peak, following the opd excursion, reveals a further difference with the bare Rh(111) behaviour, **Figure 4b, c**: the typical skewed stripping peak itself, which decays abruptly to background levels for the unobstructed Rh(111) electrode, is in the case of the nanomesh followed by a plateau where more—the current density is ca. 3 times higher—than only capacitive current flows. The length of this plateau decreases the more opd cycles are performed, and a fine structure that includes the reverse upd peak at 0.47 V develops.

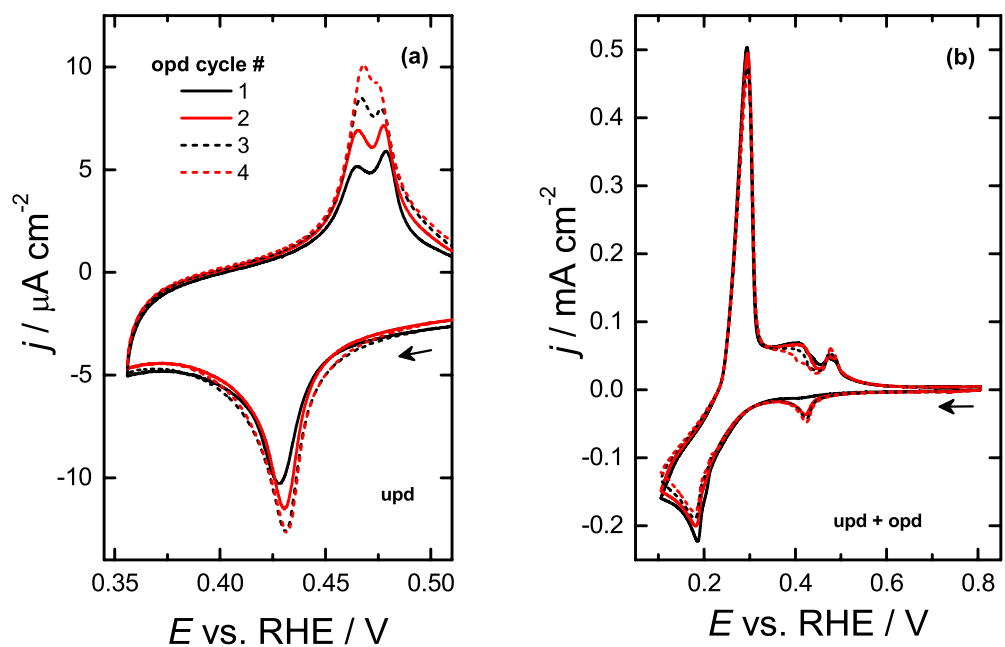

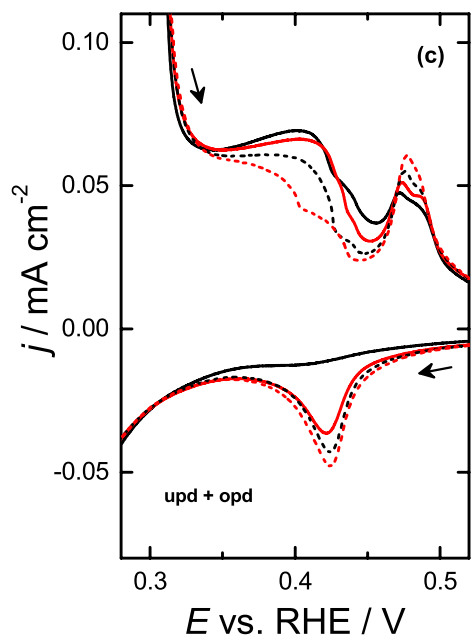

**Figure 4.** (a) Progression of upd area (inner to outer trace) as a function of the number of opd cycles as indicated for h-BN/Rh(111) (wafer sample) in 1 mM $Cu^{2+}$ + 0.1 M $H_2SO_4$. (b) Four successive cycles spanning the upd + opd region. (c) Zoom-in of Cu post-stripping peak region in panel (b). Scan rate, (a) 2 mV s$^{-1}$; (b,c) 10 mV s$^{-1}$.

Atomic force microscopy (AFM) imaging of the nanomesh surface after emersion of the sample at E = 0.25 V is shown in **Figure 5a**, in which clusters with a height of 1.8 nm can be discerned. The corrugation of the nanomesh itself (3.2 nm lattice constant) is not resolved.

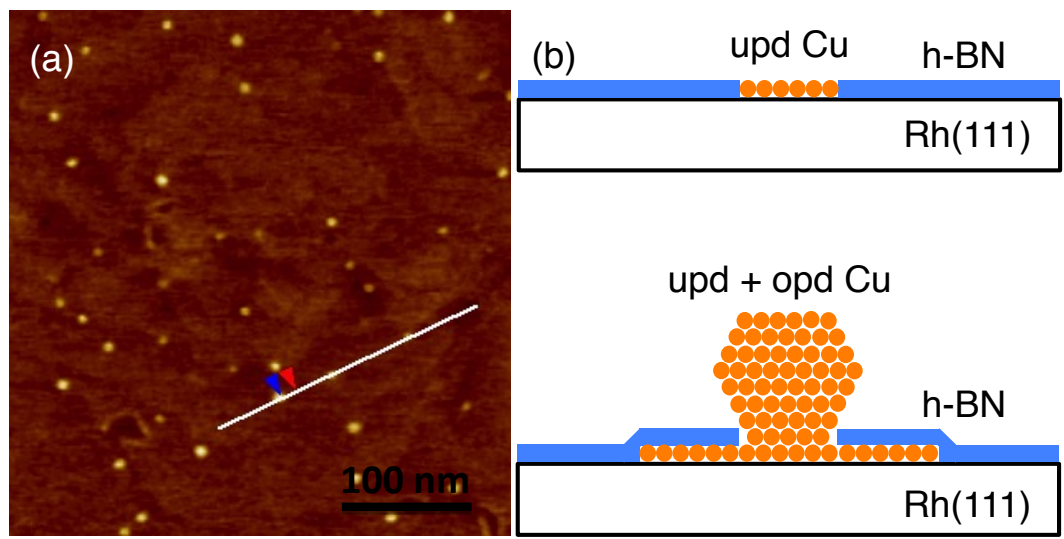

**Figure 5.** (a) AFM image of h-BN/Rh(111) (wafer sample) after Cu opd, revealing clusters on the surface. The vertical distance between the arrows along the line profile is 1.8 nm. (b) Scheme illustrating Cu upd in h-BN defects (top) and Cu cluster formation accompanied by opd-induced h-BN delamination (bottom).

Finally, the Cu upd signature was used to explore the onset of the nanomesh oxidation, as shown in **Figure 6**. To this end, the positive limit of the CV was extended to 1.25 V, and the effect on the Cu upd peaks recorded. In order to limit the dwell time of the substrate at potentials values where irreversible changes to the nanomesh take place, the scan rate during the positive excursion was set to 50 mV s$^{-1}$. As before, scans limited to the Cu upd region were performed with a scan rate of 2 mV s$^{-1}$. **Figure 6** shows Cu upd CVs after exposure of the nanomesh to one (red full trace) or two (red dashed trace) potential excursions to 1.25 V. Whereas the area under the Cu upd peak does not change significantly, the upd peak is found at more negative potentials and is markedly broader (full width at half maximum changes from 21 mV to 33 mV after a single 1.25 V excursion and to 38 mV after two). Strikingly, the double-peak shape of the reverse upd feature changes markedly following the electrochemical oxidation: the left peak decreases drastically whereas the peak at more positive potentials remains largely unchanged.

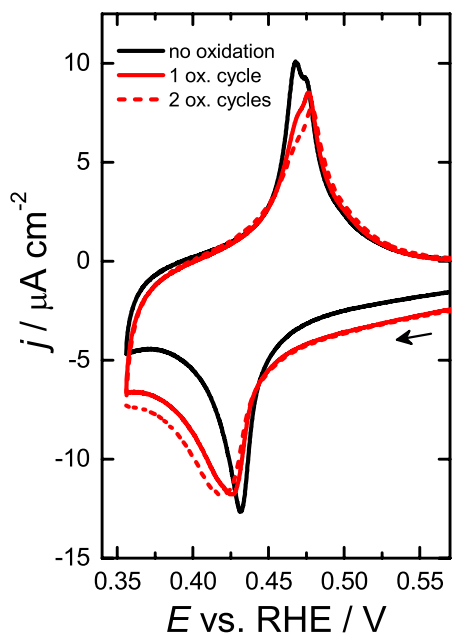

**Figure 6.** Change in Cu upd signature on h-BN/Rh(111) in 1 mM Cu$^{2+}$ + 0.1 M H$_2$SO$_4$ before and after potential excursion to +1.25 V.

## 4. Discussion

The cyclic voltammetric behaviour of Rh(111) in sulfuric acid, **Figure 1a**, matches previous reports [23-25]. If Cu$^{2+}$ is added to this solution, upd and, at more negative potentials, opd of copper is observed, **Figure 1b**. Copper upd on Rh(111) has been studied previously, not as abundantly as for instance on Au or Ag surfaces, but in sufficient detail for our purpose [23, 29]. Electrochemical scanning tunnelling microscopy studies [29] have revealed that co-adsorption of (hydrogen)sulfate yields an important contribution towards structure formation in the upd layer.

It has been argued by Anjos and co-workers that the Cu upd process starts at potentials more positive than 0.5 V, as an additional peak feature of limited intensity (estimated 10% of a monolayer) was reported by these authors [23]. Furthermore, X-ray photoelectron spectroscopy (XPS) of the surface after emersion of the electrode at potentials negative from this pre-peak confirmed the presence of copper on the surface. In the present author's opinion, these observations can be completely explained by formation of Rh surface oxide, as expected on the basis of the CV in sulfuric acid solution alone, **Figure 1a**. At potentials above 0.8 V, the formation of this surface oxide is expected to occur, and the presence of $Cu^{2+}$ in solution will not preclude this process. The presence of Cu after emersion as detected by XPS can be explained by electrostatic adsorption of $Cu^{2+}$ on the partially oxidised Rh substrate [30].

Comparison of the voltammetric behaviour of the h-BN/Rh(111) nanomesh in 0.1 M $H_2SO_4$, **Figure 2**, with that of the bare substrate, clearly indicates that the h-BN overlayer protects the underlying rhodium against electrochemical oxidation, up to potentials where the h-BN layer itself is oxidatively removed. Once this is the case, all features typical of Rh(111) voltammetry are observed. The oxidation products of h-BN likely include boron nitride oxide, which has been predicted theoretically [31], but will require further experiments (e.g. XPS following emersion and vacuum transfer of a partially oxidised nanomesh sample). The presence of the h-BN layer, on the other hand, does not preclude H upd on the underlying Rh(111). Indeed, the intercalation of hydrogen between the h-BN overlayer and its Rh(111) substrate in the boron nitride nanomesh is the crux of the electrochemical switching of its adhesive properties [15]. As the amount of intercalating H is about one-third of the amount adsorbing on bare Rh(111), this quantity cannot be used to determine the microscopic surface area of the metal. Moreover, it has been claimed that protons can pass the intact h-BN layer, as well as several other 2D materials [16], so that defects in the 2D layer cannot be conveniently studied using this process. More recently, evidence has emerged that also defective monolayer-thick membranes may display pronounced selectivity [32], based on the presence of adsorbed ions.

The results presented in **Figure 3** strongly suggest that instead, Cu upd (or by extension, that of other metals) is very promising in determining the defect area of metals covered with 2D materials. The upd process itself appears innocuous where the defects in the h-BN layer are concerned, as its

– 15 –electrochemical signature remains unchanged on multiple cycling, and holding the potential at the negative limit of the upd region for up to 600 s did not increase the reverse signal. Cu opd, on the other hand, induces clear and permanent changes in the h-BN/Rh(111) system, which, again, can be detected in a subsequent upd scan. Scanning probe (in particular EC-STM) and/or electrochemistry–UHV transfer experiments will be required to resolve the nature of the induced changes, but fall outside the scope of this report. We conjecture at this stage that the non-innocent opd process involves intercalation of the deposited metal, originating at the initial defects and propagating radially, as sketched in **Figure 5b**. A similar process has been documented by the Greber group even before the boron nitride nanomesh was discovered [33]: on depositing Co in UHV on h-BN/Ni(111), on-top cluster formation *versus* intercalation of Co 2D islands was observed as a function of the substrate temperature, and was linked to an activation barrier for intercalation of ~0.25 eV to be overcome thermally. A similar barrier in the present case may be crossed electrochemically. On stripping of the overpotentially-deposited metal, permanent changes to the h-BN/Rh(111) interface are in evidence. These may take the form of delaminated areas surrounding initial defects or even "nanoscroll" formation in the h-BN overlayer [34].

The possibility of Cu intercalation also at potentials less negative than the opd region warrants further attention. Indeed, for the closely related case of Cu upd on Au(111) covered by a thiolate self-assembled monolayer, progressive Cu deposition at the thiolate–gold interface was observed under upd conditions [35], albeit very slowly. This phenomenon was shown to initiate at incidental or intentional (e.g. STM tip-induced) defects in the thiolate layer, and to propagate radially over time. While the conclusive demonstration of the equivalent effect and the required conditions for the h-BN/Rh(111) nanomesh will require *in situ* experiments, or prolonged upd conditions followed by emersion and *ex situ* imaging, it is safe to say that the quantification of defects as proposed in this paper, because of the very different timescales involved, is not affected by this possibility.

A further consequence of possible metal intercalation under upd conditions is that the collective defect area extracted from electrochemical upd charges should be considered as an upper limit. In view of the binding mechanism of h-BN on Rh(111), with non-equivalent pore and wire regions, the microscopic situation differs depending on where the defect occurs. Even in the minimal scenario where each primary circular defect of diameter $d$ is surrounded by a ring-shaped intercalation region of the same width, a simple geometric calculation shows that in this case, the electrochemical measurement would overestimate the actual primary defect area by a factor 9. Based on the data presented in **Figure 3a**, this upper limit equals 0.7% of the geometric sample area, but may correspond to a primary defect area fraction as low as 0.08%.

The shape of the stripping (that is, the inverse opd process) and inverse upd peaks appear to contain detailed information on the microscopic structure of the modified h-BN/Rh(111) interface. The plateau next to the Cu stripping peak in **Figure 4c** indicates that complete stripping of the deposited metal requires more positive potentials than on unobstructed Rh(111). Similarly, the double peak feature during the reverse upd process (visible in Figure 3a, 4a and 6), and in particular the ratio between the two peaks, is clearly highly sensitive to the exact conditions of the experiment, and to the history of the sample. Assuming the scheme in **Figure 5b** as a working hypothesis, we propose that the peak at more negative potentials during reverse upd reflects the openly accessible defects in the h-BN/Rh(111) nanomesh: following their selective electrochemical oxidation (Figure 6), this upd peak decreases markedly whereas the peak at more positive potentials remains virtually unaffected. The second reverse upd peak, at slightly more positive potentials, may therefore reflect the confined but accessible parts of the h-BN/Rh(111) system (i.e. the areas surrounding the uncovered defect in **Figure 5b**). Along similar lines, metals upd has been used to nanoengineer thiolate self-assembled monolayers [36]. Based on our results, it appears that metals *over*potential deposition may be equally useful in nanoengineering covalent 2D networks.

From the AFM data in **Figure 5a**, it appears that opd leads to the formation of random metal clusters on the h-BN/Rh(111) surface. We consider likely that the clusters originate at the sites of initial primary defects, to which only the upd process is sensitive, as it requires exposed substrate atoms to proceed. The h-BN layer is sufficiently thin for electron tunnelling to occur, thus allowing electrochemical electron transfer [37]. Consequently, for opd, the condition of exposed substrate atoms does not exist, and in principle opd Cu may be formed on top of the intact h-BN layer through the normal nucleation and growth process. However, under the experimental conditions used here (in particular, limited excursion into the opd region), the nuclei likely remain limited to the defects that underwent Cu upd initially. Each of the nuclei thereby functions as an ultramicroelectrode and quickly depletes its surroundings of $Cu^{2+}$. The cluster distribution visible in **Figure 5a** indicates that for this particular sample, random point defects rather than line defects originating at domain boundaries during high-temperature growth of the h-BN layer [20] appear to be dominant. In view of the strong covalent network represented by the h-BN layer, we consider it unlikely that the number density of defects would change under these mild electrochemical conditions.

5. **Conclusions**

In 0.1 M sulfuric acid, the h-BN/Rh(111) nanomesh has an electrochemical window of ca. 1 V, which is larger than its substrate's, and is limited by hydrogen evolution on the negative side and by h-BN oxidative removal at positive potentials. Copper upd on h-BN/Rh(111) as a nondestructive method can be used successfully to determine the collective defect area in the h-BN layer, arising for instance from line defects during CVD growth. At more negative potentials, opd of copper causes permanent changes to the boron nitride nanomesh, the quantification of which can again be carried out by measuring upd. AFM confirms metal cluster growth during opd; we hypothesise that this growth is accompanied by non-innocent Cu adsorption on Rh(111), involving intercalation and irreversible delamination around initial defects in the h-BN layer. We propose that metal upd may be used universally to study defects in 2D materials on metal substrates in the presence of suitable metal ions in solution.

**Acknowledgements**

Financial support by the Swiss National Science Foundation within the funding instrument "Sinergia", the European Union through FP7 Marie Curie European reintegration grant "Templates" and Austrian Science Fund (FWF) project I3256-N36 is gratefully acknowledged. The author thanks Thomas Greber (University of Zürich) for discussions and nanomesh samples, and Willem Vanderlinden (KU Leuven) for AFM measurements.